\documentclass[prd,tightenlines,showpacs,nofootinbib,eqsecnum,amsfonts,amssymb,amsmath,twocolumn,preprintnumbers]{revtex4}

\usepackage[english]{babel}
\usepackage{inputenc}
\usepackage{graphicx}
\usepackage{mathrsfs}
\usepackage{epsfig,bbm}
\usepackage{bm}
\usepackage{color}
\usepackage{tensor}
\usepackage{aeguill}
\usepackage{microtype}
\usepackage{xspace}
\usepackage{wasysym}
\usepackage{amsthm}
\usepackage{hyperref}

\newcommand\ph{\ensuremath{\varphi}}
\newcommand\eps{\ensuremath{\varepsilon}}

\newcommand\define{\overset{\text{def.}}{=}}
\newcommand\vect[1]{\boldsymbol{#1}}
\newcommand\ex[1]{\mathrm{e}^{#1}}
\renewcommand\i{\ensuremath{\mathrm{i}}}

\newcommand{\tr}{\ensuremath\mathrm{Tr}}

\newcommand\e[1]{_{\text{#1}}}
\newcommand\h[1]{^{\text{#1}}}

\newcommand\dd{ \mathrm{d} }
\newcommand{\Dd}{\mathrm{D}}
\newcommand{\ddf}[3][]{\frac{\dd^{#1} #2}{\dd {#3}^{#1}}}

\newcommand\pa[1]{\left( #1 \right)}
\newcommand\pac[1]{\left[ #1 \right]}

\newcommand{\optical}{\mathcal{R}}
\newcommand{\jacobi}{\mathcal{D}}

\newcommand{\evolution}{\mathcal{E}}

\newcommand{\intmatrix}{\mathcal{I}}

\newcommand{\B}{\mathcal{B}}

\newcommand{\omegac}{\tilde{\omega}}
\newcommand{\conformal}[1]{\tilde{#1}}

\newcommand{\source}{\e{s}}
\newcommand{\obs}{\e{o}}
\newcommand{\jacobic}{\conformal{\jacobi}}
\newcommand{\opticalc}{\conformal{\optical}}

\renewcommand{\define}{\equiv}

\begin{document}

\title{Light propagation in a homogeneous and anisotropic universe}

\author{Pierre Fleury}
\email{fleury@iap.fr}

\author{Cyril Pitrou}
\email{pitrou@iap.fr} 

\author{Jean-Philippe Uzan}
\email{uzan@iap.fr}             
 
\affiliation{
Institut d'Astrophysique de Paris, UMR-7095 du CNRS, Universit\'e
Pierre et Marie Curie, \\
98 bis bd Arago, 75014 Paris, France, and\\
Sorbonne Universit\'es, Institut Lagrange de Paris, 98 bis bd Arago, 75014 Paris, France.\\
}

\begin{abstract}
This article proposes a comprehensive analysis of light propagation in an anisotropic and spatially homogeneous Bianchi~I universe. After recalling that null geodesics are easily determined in such a spacetime, we derive the expressions of the redshift and direction drifts of light sources; by solving analytically the Sachs equation, we then obtain an explicit expression of the Jacobi matrix describing the propagation of narrow light beams. As a by-product, we recover the old formula by Saunders for the angular diameter distance in a Bianchi~I spacetime, but our derivation goes further since it also provides the optical shear and rotation. These results pave the way to the analysis of both supernovae data and weak lensing by the large-scale structure in Bianchi universes. 
\end{abstract}

\date{\today}
\pacs{98.80}

\maketitle

\section{Introduction}\label{sec1}

The standard cosmological model relies heavily on the assumption that on the large scale it is well described by a spacetime with homogeneous and isotropic spatial sections. All cosmological observations tend to agree with this geometrical assumption, and to back up the predictions of the $\Lambda$CDM model with a primordial inflationary phase. 

A lot of efforts are invested in order to determine whether the source of the acceleration of the expansion of the Universe is due to a cosmological constant or has a dynamical origin (new matter fields dubbed dark energy or gravity beyond general relativity); see e.g. Refs.~\cite{Uzan:2006mf,UzanDE}. It has also revived the importance of testing the validity of the Copernican principle.

While a primordial shear decays if it is not sourced, late-time anisotropy appears in many phenomenological models of dark energy~\cite{Bucher:1998mh,mota1,mota3,Appleby2,Appleby:2009za} and is a generic prediction of bigravity models~\cite{bigrav} and backreaction~\cite{marozzi}. Contrary to the former~\cite{PPU1,PPU2,cmbaniso,Battye:2009ze,Battye:2005mm}, the latter remains weakly constrained by the observation of the cosmic microwave background temperature field; this naturally stimulated analyses based, e.g., on the observation of supernovae~\cite{Mota2,schucker_bianchi_2014,Jimenez:2014jma,2001MNRAS.323..859K,2011MNRAS.414..264C,2013A&A...553A..56K,2010JCAP...10..018B,2013PhRvD..87l3522C,2007A&A...474..717S,2010JCAP...12..012A,2013A&A...560A..90F,2014JCAP...03..007A,2014arXiv1410.5562A}, or using low-redshift galaxies~\cite{Appleby,2014MNRAS.445L..60Y}. Besides the strict detection of anisotropy, drawing quantitative conclusions from such analyses requires one to understand how light propagates through an anisotropic universe. This issue has been addressed since the late sixties~\cite{saunders_observations_1968,saunders_observations_1969,1968PThPh..40..264T,1970CMaPh..19...31M}, in particular, a remarkably simple expression of the angular diameter distance in Bianchi~I models was found by Saunders~\cite{saunders_observations_1968,saunders_observations_1969} using observational coordinates~\cite{obscoord}, and recently rederived in Ref.~\cite{schucker_bianchi_2014}.

The purpose of this article is to provide a complete analytical study of light propagation in Bianchi~I spacetimes. On the one hand, the integration of the null geodesic equation (though already well known) allows us to derive the expressions of the redshift, redshift drift and position drift of an arbitrary light source. More importantly, on the other hand, we solve the Sachs equation governing the geometry of geodesic bundles. From the resulting Jacobi matrix, we not only recover Saunders' formula for the angular diameter distance, but also characterize the whole lensing properties generated by anisotropy. These results pave the way to the computation of the lensing $B$-mode signal induced in an anisotropic universe---as predicted in Ref.~\cite{Bmodes}---since it provides the background result for the general computation in perturbed Bianchi models.

The article is organized as follows. After summarizing the main geometrical properties of a Bianchi~I universe in Sec.~\ref{sec:Bianchi_I_spacetime}, and the laws of geometric optics in curved spacetime in Sec.~\ref{sec3}, we solve the null geodesic equation and derive the expressions of the redshift and direction drifts in Sec.~\ref{sec:light_rays_BIanchi_I}. One technical key point of our construction is the use of a conformal transformation, whose dictionary is detailed in Sec.~\ref{sec:dictionary}. The heart of our derivation is then exposed in Secs.~\ref{sec:Sachs_basis} and \ref{sec:Jacobi matrix}, in which, respectively, we construct the Sachs basis and obtain the expression of the Jacobi matrix---see in particular Eq.~(\ref{eq:Jacobi_abtract}). An algorithmic way of using our results is proposed in Sec.~\ref{sec:summary}. Finally, in the Appendix we give a proof of the result of Ref.~\cite{saunders_observations_1969}.

\section{The Bianchi~I spacetime}\label{sec:Bianchi_I_spacetime}

The classification of spatially anisotropic and homogeneous spacetimes~\cite{EllisAndGolum} is based on the Bianchi's classification of
homogeneous but not necessarily isotropic three-dimensional spaces~\cite{LuigiBianchi}. The spatial sections of these spacetimes are Bianchi spaces characterized by their Riemann tensor (more precisely the Riemann tensor of the induced metric on the spatial sections), and the full geometry is then determined from the extrinsic curvature of the spatial sections. The simplest of these spacetimes is Bianchi~I, which enjoys Euclidean spatial sections, that is with a vanishing Riemann tensor of the induced 3-metric. Its metric reads simply
\begin{equation}\label{e.bianchi1}
 \dd s^2 = g_{\mu\nu} \dd x^\mu \dd x^\nu = -\dd t^2 + a^2(t) \gamma_{ij} \dd x^i \dd x^j,
\end{equation}
where the spatial metric is given by
\begin{equation}\label{e.defgamma}
 \gamma_{ij} = \ex{2\beta_i(t)} \delta_{ij}
\end{equation} 
with the constraint 
\begin{equation}\label{e.contrainte}
 \sum_{i=1}^3 \beta_i=0.
\end{equation}
The inverse spatial metric is $\gamma^{ij} = \ex{-2\beta_i(t)} \delta_{ij}$, such that $\gamma_{ik}\gamma^{kj} =\delta_i^j$. With this choice of the metric parametrization, the volume expansion is encoded in the scale factor $a(t)$, while the evolution of $\gamma_{ij}$ is volume preserving, thanks to the condition~(\ref{e.contrainte}). The conformal time~$\eta$ is defined from cosmic time~$t$ by the usual relation $\dd t = a \dd\eta$.

The \emph{conformal} shear (rate) tensor~$\sigma_{ij}$ is defined by
\begin{equation}
 \sigma_{ij} \define \frac{1}{2} (\gamma_{ij})' = \beta_i' \gamma_{ij}
\end{equation}
where a prime denotes a derivative with respect to conformal time~$\eta$. Its geometrical interpretation is simple as it is directly related to the traceless part of the extrinsic curvature of space sections, whose components are just $a^2 \sigma_{ij}$. The indices of $\sigma_{ij}$ are respectively raised and lowered by $\gamma^{ij}$ and $\gamma_{ij}$. Note that $\gamma_{ik} \gamma^{kj} = \delta_i^j$ implies
\begin{equation}
 \sigma^{ij}  = \beta_i' \gamma^{ij} = -\frac{1}{2}(\gamma^{ij})'\,,
\qquad \sigma_i^j = \beta_i' \delta_i^j\,.
\end{equation}

Since the spatial sections are homogeneous, there exists a class of preferred observers---called {\em fundamental observers}---for which space indeed looks homogeneous. They are comoving with respect to the Cartesian coordinate system introduced in Eq.~(\ref{e.bianchi1}), and cosmic time~$t$ represents their proper time, so that the four-velocity of fundamental observers reads $u^\mu= (\partial_t)^\mu$.

For a universe filled by a homogeneous fluid, the stress-energy tensor is
\begin{equation}
 T_{\mu\nu}  = \rho u_\mu u_\nu + P(g_{\mu\nu}+u_\mu u_\nu) + \Pi_{\mu\nu},
\end{equation}
with $\rho$ and $P$ being the energy density and the isotropic pressure, and where $\Pi_{\mu\nu}$ the anisotropic stress. This latter symmetric tensor is traceless and spatial, in the sense that $u^\mu\Pi_{\mu\nu}=0=\Pi^\mu_\mu$.  We further define the \emph{conformal} anisotropic pressure by $\pi_{ij}\define \Pi_{ij}/a^2$ and $\pi^{ij}\define \Pi^{ij} a^2$ such that the indices of $\pi_{ij}$ are respectively raised and lowered by
$\gamma_{ij}$ and $\gamma^{ij}$, as is the case for $\sigma_{ij}$. 

The Einstein field equations then read
\begin{align}
\mathcal{H}^2 &= \frac{8\pi G}{3} a^2\rho + \frac{\sigma^2}{6},  \\
\mathcal{H}' &= -\frac{4\pi G}{3} a^2(\rho+3P) - \frac{\sigma^2}{3}, \\
(\sigma^i_j)' &= - 2\mathcal{H} \sigma^i_j + 8\pi G a^2 \pi^i_j,
\label{eq:dynamics_shear}
\end{align}
where $\mathcal{H}\define a'/a$ is the conformal expansion rate, and 
\begin{equation}
 \sigma^2 \define \sigma^{ij}\sigma_{ij} = \sum_{i=1}^3 \pa{\beta_i'}^2.
\end{equation}

\section{Geometric optics in a general curved spacetime}\label{sec3}

This section briefly reviews the essential equations governing light propagation in curved spacetime, its main purpose being to fix the notations. For further details, we refer the reader e.g., to the textbook~\cite{1992grle.book.....S} or the review~\cite{2004LRR.....7....9P} and our previous papers~\cite{FDU1,FDU2,Fleury:2014gha}.

\subsection{Light rays}

Electromagnetic waves, described by Maxwell electrodynamics and identified to light rays in the eikonal approximation, are shown to follow null geodesics~\cite{1973grav.book.....M}. If $v$ denotes an affine parameter along such a geodesic, its tangent vector $k^\mu=\dd x^\mu/\dd v$---which is also the wave four-vector of the electromagnetic signal---is a null vector ($k^\mu k_\mu=0$) that satisfies the geodesic equation
\begin{equation}\label{eq:null_geodesic_equation}
 \frac{\Dd k^\mu}{\dd v} \define k^\nu \nabla_\nu k^\mu = 0,
\end{equation}
where $\nabla_\mu$ denotes the covariant derivative associated to the metric $g_{\mu\nu}$.

An observer whose worldline intersects the ray can naturally define the notions of pulsation (or energy)~$\omega$, and spatial direction of observation~$d^\mu$, by performing a $3+1$ decomposition of $k^\mu$ with respect to his own four-velocity~$u^\mu$, as
\begin{equation}\label{eq:3+1}
k^\mu = \omega (u^\mu-d^\mu),
\end{equation}
where $\omega\define -u_\mu k^\mu$, and $d^\mu$ is a unit spatial vector, i.e.  $u^\mu d_\mu=0$ and $d^\mu d_\mu=1$.

\subsection{Light beams}

A (narrow) light beam is a collection of neighboring light rays, i.e. an infinitesimal bundle of null geodesics. The behavior of any such geodesic, with respect to an arbitrary reference one, is described by the separation (or connecting) vector $\xi^\mu$. If all the rays converge at a given event $O$---the observation event ``here and now'' denoted with the index ``o'' in the following---then $\xi^\mu(v\obs)=0$. The evolution of $\xi^\mu(v)$ along the beam is governed by the geodesic deviation equation~\cite{1973grav.book.....M}
\begin{equation} \label{eq:GDE}
  \frac{\Dd^2 \xi^\mu}{\dd v^2} = R\indices{^\mu_\nu_\rho_\sigma} k^\nu k^\rho \xi^\sigma,
\end{equation}
where $R_{\mu\nu\rho\sigma}$ denotes the Riemann tensor.

\subsection{Sachs basis}

For any observer whose worldline intersects the light beam at an event different from $O$, the beam has a non-zero extension since \textit{a priori} $\xi^\mu \not= 0$. The observer can thus project it on a \emph{screen} to characterize its size and shape. This screen is by essence a two-dimensional spacelike plane chosen to be orthogonal to the local line-of-sight $d^\mu$. Thus, if $(s_A^\mu)_{A=1,2}$ is an orthonormal basis of the screen, then
\begin{equation}\label{eq:Sachs_orthogonality}
s_A^\mu u_\mu=s_A^\mu d_\mu=0,
\qquad
s_A^\mu s_{B\mu} =\delta_{AB}.
\end{equation}
Note that, by virtue of Eq.~\eqref{eq:3+1}, we also have $s_A^\mu k_\mu=0$.

Now, consider a flow of observers lying all along the beam [defining a four-velocity field $u^\mu(v)$] who want to compare the size, shape, and orientation of the pattern they observe on their respective screen. To avoid any spurious rotation of this pattern, one has to further impose that the basis vectors $(s_A^\mu)_{A=1,2}$ are Fermi-Walker transported along the beam,
\begin{equation}\label{eq:Fermi_Walker_general}
  S^\mu_\nu \frac{\Dd s_A^\nu}{\dd v} = 0,
\end{equation}
where 
\begin{equation}\label{screendef}
S^{\mu\nu}\define\delta^{AB} s_A^\mu s_B^\nu = g^{\mu\nu}+u^\mu u^\nu-d^\mu d^\nu
\end{equation} 
is the screen projector. The transport rule \eqref{eq:Fermi_Walker_general} must be understood as: $s_A^\mu$ is parallel transported as much as possible while keeping it orthogonal to $u^\mu$ and $d^\mu$.

The set of vectors $(s_A^\mu)_{A=1,2}$ satisfying Eqs.~\eqref{eq:Sachs_orthogonality} and \eqref{eq:Fermi_Walker_general} is known as the \emph{Sachs basis}.

\subsection{Jacobi matrix}

The screen projection of the connecting vector, $\xi_A\define s_A^\mu\xi_\mu$, represents the relative position on the screen of the two light spots associated with two rays separated by $\xi^\mu$. Similarly, and if we set by convention $\omega\obs=1$, $\theta_A \define - (\dd \xi_A/\dd v)\obs$ represents the angular separation of those rays, as observed from $O$. The matrix relating $\xi_A(v)$ to $\theta_A$ via
\begin{equation}\label{eq:Jacobi_def}
  \xi_A(v) = \jacobi_{AB}(v\leftarrow v\obs) \theta_B,
\end{equation}
is known as the \emph{Jacobi matrix}. The equation governing its evolution along the beam derives from the geodesic deviation equation \eqref{eq:GDE}, and reads
\begin{equation}\label{eq:Jacobi_matrix_equation}
  \ddf[2]{\jacobi_{AB}}{v} = \optical_{AC} \jacobi_{CB},
\end{equation}
where $\optical_{AB}=-R_{\mu\nu\rho\sigma}s_A^\mu k^\nu s_B^\rho k^\sigma$ is called the \emph{optical tidal matrix}. Note that the position of the screen indices $A, B, C,\ldots$ does not matter, since they are raised and lowered by $\delta_{AB}$. The initial conditions for Eq.~\eqref{eq:Jacobi_matrix_equation} are
\begin{align}
\jacobi_{AB}(v\obs\leftarrow v\obs) &= 0,
\label{eq:initial_Jacobi_1}\\
\ddf{\jacobi_{AB}}{v}(v\obs\leftarrow v\obs) &= -\delta_{AB}.
\label{eq:initial_Jacobi_2}
\end{align}

By definition~\eqref{eq:Jacobi_def}, the Jacobi matrix relates the size and shape of the beam to its observed angular aperture. It is thus naturally related to the angular diameter distance $D\e{A}$, linked to the ratio of the area~$\dd^2 A\source$ of a (small) light source to its observed angular size~$\dd^2\Omega\obs$,
\begin{equation}\label{eq:angular_distance_Jacobi}
  D\e{A} \define \sqrt{\frac{\dd^2 A\source}{\dd^2 \Omega\obs}} = \sqrt{\det \vect{\jacobi}(v\source\leftarrow v\obs)}.
\end{equation}

More generally, the Jacobi matrix encodes all the information about the deformation of a light beam with its propagation through a curved spacetime, i.e., gravitational lensing. A canonical decomposition\footnote{Although the authors have never seen this decomposition used in the literature so far, they advocate that it is more meaningful than the standard one
\begin{equation}
 \vect{\jacobi} = D\e{A}\h{FL}
 \begin{bmatrix}\label{eq:usual_decomposition_Jacobi}
 1 - \kappa - \gamma_1 & \gamma_2 - \omega \\
 \gamma_2 + \omega & 1 - \kappa + \gamma_1
 \end{bmatrix},
\end{equation}
which explicitly makes use of the angular distance in a Friedmann-Lema\^itre (FL) spacetime, $D\e{A}\h{FL}$, and the ``convergence''~$\kappa$, ``shear''~$\gamma_{1,2}$, and ``rotation''~$\omega$ with respect to it. Additionally to the fact that such a decomposition relies on the choice of a specific background (namely FL), the quantities $\kappa$, $\gamma_{1,2}$, and $\omega$ lose their geometrical meaning for \emph{finite} (noninfinitesimal) lensing effects. This is why, for instance, $\gamma$ appears in the expression of the magnification. It is not the case for the decomposition proposed in Eq.~\eqref{eq:decomposition_Jacobi}.
}
of $\vect{\jacobi}$ that makes such effects explicit is
\begin{equation}\label{eq:decomposition_Jacobi}
\vect{\jacobi}
=
D\e{A}
\underbrace{
\begin{bmatrix}
\cos\psi & \sin\psi \\
-\sin\psi & \cos\psi
\end{bmatrix}
}_\text{rotation}
\underbrace{
\exp
\begin{bmatrix}
-\Gamma_1 & \Gamma_2 \\
\Gamma_2 & \Gamma_1
\end{bmatrix}
}_\text{shear}.
\end{equation}
According to this decomposition, the real size and shape of a light source is obtained from its image by performing the following transformations: (i) an area-preserving shear, (ii) a global rotation, (iii) a global scaling.

\section{Geodesic motion in Bianchi I}\label{sec:light_rays_BIanchi_I}

There is a simple and elegant way to determine geodesics in a spacetime with spatial homogeneity, without explicitly solving the geodesic equation~\eqref{eq:null_geodesic_equation}. It relies on the basic fact~\cite{dubook} that for any Killing vector $\zeta^\mu$ of the metric, the scalar $k^\mu\zeta_\mu$ is constant along the geodesic whose tangent vector is $k^\mu$ (whether it is null or not).

\subsection{Light rays}

Since $\partial_i$ is a Killing vector of the Bianchi~I spacetime, the quantity $g(\partial_i,k)=k_i$ is a constant of geodesic motion. Moreover, since $k$ is a null vector, $\omega^2\define (k^t)^2=g^{ij} k_i k_j$ and the wave four-vector thus reads
\begin{align}\label{e41}
 k_i &= \mathrm{cst},	\qquad  k^i = a^{-2} \gamma^{ij} k_j \not=\mathrm{cst},\\
\omega &= \frac{\omegac}{a},
\end{align}
where
\begin{equation}\label{Defomegatilde}
\omegac \define \sqrt{\sum_{i=1}^3 (\ex{-\beta_i} k_i)^2}.
\end{equation}
The components of the direction of observation vector $d^\mu$ are, by definition,
\begin{equation}\label{di}
d_i = -k_i/\omega\,,\qquad d^i = -k^i/\omega\,.
\end{equation}
From now on, we set by convention $a(t\obs)=1$ and $\beta_i(t\obs)=0$ at $O$ ($t=t\obs$), hence the redshift is given by
\begin{equation}\label{e.defz}
1+z \define \frac{\omega\source}{\omega\obs} = \frac{1}{a(t\source)} \sqrt{\sum_{i=1}^3 \pac{\ex{-\beta_i(t\source)} k_i}^2}.
\end{equation}

The constants of motion $k_i$ are directly related to the direction in which the observer at $O$ needs to look to detect the light signal. Indeed, with the conventions specified above, at the observation event $(g_{ij})\obs=\delta_{ij}$, moreover we have used the remaining freedom to set $\omega\e{o}=1$, so that $-k_i=(d_i)\obs$ is a unitary Euclidean three-vector.

\subsection{Parentheses: On timelike geodesics}

The previous reasonings also apply to timelike geodesics. Consider a general observer, whose four-velocity $v^\mu$ can be decomposed with respect to the four-velocity~$u^\mu$ of the fundamental (comoving) observers as
\begin{equation}
v^\mu = E u^\mu +p^\mu
\end{equation}
with $u^\mu p_\mu=0$ and $u^\mu u_\mu=-1$. Since $v^\mu v_\mu=-1$, we have $p^\mu p_\mu=-1+E^2$. Now the constancy of $p_i$ implies that $E^2=1+a^{-2}\gamma^{ij}p_ip_j \rightarrow 1$ as $t$ increases (in an expanding universe), so that the worldline of the observer tends to align with the worldline of the fundamental observers, i.e. the Hubble flow, exactly as in Friedmann-Lema\^{\i}tre spacetimes~\cite{pubook}.

\subsection{Redshift and direction drifts}

\subsubsection{Redshift drift}

As originally pointed out by Sandage and McVittie~\cite{sandage,sandage2} a consequence of the expansion of the Universe is the existence of a drift of the cosmological redshifts. This effect is thought to be observationally accessible~\cite{codex,codex2} in the standard cosmological framework~\cite{zdot,zdotuce,realtimecosmo,realtimecosmo2}.

Consider a photon received at $t\obs+\delta t\obs$, corresponding to the emission time~$t\source+\delta t\source$; by definition of the redshift,
\begin{equation}
1+z+\delta z
\define
\frac{\omega(t\source+\delta t\source)}{\omega(t\obs+\delta t\obs)}
=
\sqrt{ \frac{g^{ij}(t\source+\delta t\source) k_i k_j }
					{g^{ij}(t\obs+\delta t\obs)k_i k_j}}.
\end{equation}
We can expand the above formula at first order in $\delta t\obs$ and $\delta t\e{s}$ using $g^{ij}=\gamma^{ij}/a^2$, which leads to
\begin{equation}
\frac{\delta z}{1+z} = \frac{\delta t\obs}{a\obs} \pa{ \mathcal{H} + \sigma^i_j d_i d^j }\obs
			-\frac{\delta t\source}{a\source} \pa{ \mathcal{H} + \sigma^i_j d_i d^j }\source.
\end{equation}
Since moreover $\delta t\source/\delta t\obs = 1/(1+z)$, we finally get the redshift drift~$\dot z\obs\define\delta z/\delta t\obs$ observed by $O$:
\begin{equation}
\dot{z}\obs = (1+z) H^\parallel\obs - H^\parallel\source,
\label{eq:redshift_drift}
\end{equation}
where
\begin{equation}
H^\parallel(z,d^i) \define \frac{1}{a} \pa{ \mathcal{H} + \sigma^i_j d_i d^j }.
\end{equation}
It is interesting to notice that Eq.~\eqref{eq:redshift_drift} is identical to the one obtained in a Lema\^{\i}tre-Tolman-Bondi universe~\cite{zdotuce}, and indeed reduces to the Sandage formula~\cite{sandage,sandage2} in the isotropic case.

\subsubsection{Direction drift}

A consequence of anisotropic expansion is that, besides redshift drift, the position of a comoving light source on the observer's celestial sphere also changes with time. Let us compute the velocity of this direction drift. The position~$x^i\source$ of the source is obtained by integrating the wave vector~$k^i$ with respect to the affine parameter,
\begin{equation}
x^i\source = \int_{v\obs}^{v\source} k^i \;\dd v = \pa{ \int_{v\obs}^{v\source} a^{-2} \ex{-2\beta_i} \; \dd v } d^i\obs,
\end{equation}
where we used $k_i=\mathrm{cst}=d^i\obs$. Like for redshift drift, we can evaluate the above relation at a later observation time $t\obs+\delta t\obs$ corresponding to an emission time $t\source+\delta t\source$. If the source is comoving, then $x^i\source$ remains unchanged, so that
\begin{multline}
\pa{ \int_{v(t\obs)}^{v(t\source)} a^{-2}\ex{-2\beta_i} \; \dd v } d^i\obs \\
= \pa{ \int_{v(t\obs+\delta t\obs)}^{v(t\source+\delta t\source)} a^{-2}\ex{-2\beta_i} \; \dd v } (d^i\obs+\delta d^i\obs).
\label{eq:direction_drift_calculus}
\end{multline}
The direction drift velocity~$\dot{d}^i\obs\define \delta d^i\obs/\delta t\obs$ is finally obtained by performing a first-order expansion of Eq.~\eqref{eq:direction_drift_calculus}, using in particular $v(t+\delta t)=v(t)+\delta t/\omega$, and the result is
\begin{equation}
\dot{d}^i\obs = \pa{ \int_{v\obs}^{v\source} a^{-2}\ex{-2\beta_i} \; \dd v }^{-1} \pa{ d^i\obs - \frac{d^i\source}{1+z}}.
\label{eq:direction_drift}
\end{equation}

\section{The conformal dictionary}
\label{sec:dictionary}

The determination of the Jacobi matrix in a Bianchi~I spacetime is greatly simplified by using the fact that two conformal spacetimes have the same light cone.\footnote{In four dimensions, this result can be related to the conformal invariance of Maxwell theory. However, this property of the null geodesics holds even in higher dimensions whilst Maxwell theory is no more conformal invariant. From the physical point of view, this is due to the fact that \emph{in the eikonal approximation} all the terms which are not conformally invariant are subdominant. It follows that geometric optics enjoys more symmetries that the microscopic theory it derives from.} Let the conformal metric~$\conformal{g}_{\mu\nu}$ be defined by
\begin{equation}
g_{\mu\nu} = a^2 \conformal{g}_{\mu\nu}.
\end{equation}
\emph{Property.}---Any null geodesic for $g_{\mu\nu}$, affinely parametrized by $v$, is also a null geodesic for $\conformal{g}_{\mu\nu}$, affinely parametrized by $\conformal{v}$ with $\dd v = a^2 \dd\conformal{v}$. The associated wave four-vectors then read $\conformal{k}^\mu = a^2 k^\mu$.

As a consequence, the \emph{covariant} components of $k$ are unchanged by the conformal transformation, indeed
\begin{equation}
\tilde{k}_\mu = \tilde{g}_{\mu\nu} \tilde{k}^\nu = a^{-2} g_{\mu\nu} a^2 k^\mu = k_\mu.
\end{equation}
The four-velocities of comoving observers for both geometries are respectively $u=\partial_t$ and $\conformal{u}=\partial_\eta$, so that $\conformal{u}^\mu = a\,u^\mu$, thus
\begin{equation}
\omega \define g_{\mu\nu} u^\mu k^\nu 
				= a^{-1} \conformal{g}_{\mu\nu} \conformal{u}^\mu \conformal{k}^\nu 
				= \omegac/a.
\end{equation}
The 3+1 decomposition of $\conformal{k}^\mu$ is therefore
\begin{equation}
\conformal{k}^\mu = \omegac (\conformal{u}^\mu-\conformal{d}^\mu)
\end{equation}
with $\conformal{d}^\mu\define a\,d^\mu$ implying $\tilde d_\mu =
d_\mu/a$.

The Sachs basis~$(\conformal{s}_A^\mu)_{A=1,2}$ for the conformal geometry is related to the original one by
\begin{equation}
\conformal{s}_A^\mu = a \, s_A^\mu.
\label{eq:conformal_Sachs}
\end{equation}
One can indeed check that, with Eq.~\eqref{eq:conformal_Sachs}, the usual orthonormality and Fermi-Walker
transport conditions are preserved by the conformal
transformation,\footnote{The connections $\tilde \nabla$ and $\nabla$
are related by
\begin{equation}
\tilde \nabla_\mu V_\nu = \nabla_\mu V_\nu-V_\alpha \left[2 \delta^\alpha_{(\mu} \nabla_{\nu)} \ln a - g_{\mu\nu}g^{\alpha \beta}\nabla_\beta \ln a \right].
\end{equation}
} i.e.
\begin{equation}
\left\{
\begin{aligned}
s_A^\mu u_\mu &= 0, \\
s_A^\mu d_\mu &= 0, \\
s_A^\mu s_{B\mu} &= \delta_{AB}, \\
S^\nu_\mu k^\rho \nabla_\rho s_A^\nu &=0.
\end{aligned}
\right.
\Longleftrightarrow
\left\{
\begin{aligned}
\conformal{s}_A^\mu \conformal{u}_\mu &= 0, \\
\conformal{s}_A^\mu \conformal{d}_\mu &= 0, \\
\conformal{s}_A^\mu \conformal{s}_{B\mu} &= \delta_{AB}, \\
\conformal{S}^\nu_\mu \conformal{k}^\rho \conformal{\nabla}_\rho \conformal{s}_A^\nu &=0.
\end{aligned}
\right.
\end{equation}
In these relations, $S_{\mu\nu}$ is the screen projector defined in Eq.~(\ref{screendef}) and we have an analogous definition for the conformal geometry, which implies $S_{\mu\nu}=a^2\conformal{S}_{\mu\nu}$.

The separation four-vector~$\xi^\mu$ between two neighboring geodesics is defined by comparing events only, independently from any metric. It is therefore invariant under conformal transformations. However, its projection over the Sachs basis changes (since the Sachs basis itself changes), indeed
\begin{equation}
\xi_A \define s_A^\mu \xi_\mu = a^{-1} \conformal{s}_A^\mu a^2\conformal{\xi}_\mu = a\,\conformal{\xi}_A.
\end{equation}
The above relation allows us to relate the Jacobi matrices calculated in both geometries, and the result is
\begin{align}
\jacobi_{AB}(\mathrm{s}\leftarrow\mathrm{o}) = a\e{s} \conformal{\jacobi}_{AB}(\mathrm{s}\leftarrow\mathrm{o}),
\end{align}
which, by virtue of Eq.~\eqref{eq:angular_distance_Jacobi}, implies
\begin{equation}
D\e{A} = a\source \conformal{D}\e{A}.
\end{equation}

\section{Sachs basis in a conformal Bianchi I geometry}
\label{sec:Sachs_basis}

\emph{Important remark.}---In this section, all the calculations are performed in the conformal geometry $\conformal{g}_{\mu\nu}$. Since only intermediary results are at stake, \emph{we temporarily drop all the tildes on the vectors $\conformal{d}^\mu$, $\conformal{s}_A^\mu$ to alleviate notation}. However, we do not drop the tilde on $\omegac$ because it could lead to ambiguities.\\

By definition, the Sachs basis is purely spatial, so
\begin{equation}
 u_\mu s_A^\mu = 0.
\end{equation}
The evolution of the nonzero spatial part of $s_A^\mu$ follows from
the Fermi-Walker transport~\eqref{eq:Fermi_Walker_general}, which takes the form
\begin{equation}\label{eq:Fermi_Walker_Bianchi}
(s_A^i)' + S^i_j \sigma^j_k s_A^k = 0, 
\end{equation}
where $S^i_j=\delta^i_j - d^i d_j$ (since $u^i=0$) and we used that the only nonvanishing Christoffel coefficients are
\begin{equation}
\conformal{\Gamma}\indices{^i_0_j} = \sigma^i_j,
\qquad
\conformal{\Gamma}\indices{^0_i_j} = \sigma_{ij}.
\end{equation}
%

\subsection{General solution of the transport equation}
\label{subsec:transport_Sachs}

Let $(n^\mu_A)_{A=1,2}$ be an arbitrary orthonormal basis of the screen space (i.e. orthogonal to both $u^\mu$ and $d^\mu$), not necessarily Fermi-Walker transported along the light beam. Explicit examples of such a basis will be given in Sec.~\ref{subsec:explicit_examples}. The Sachs basis~$(s^\mu_A)_{A=1,2}$ being also an orthonormal basis of the same space, the two basis are related by a rotation
\begin{align}\label{eq:s1}
\left\{
\begin{aligned}
s_1^\mu &= \cos\vartheta \, n_1^\mu + \sin\vartheta \, n_2^\mu,\\
s_2^\mu &= -\sin\vartheta \, n_1^\mu + \cos\vartheta \, n_2^\mu.
\end{aligned}
\right.
\end{align}
Hence, provided the basis $(n^\mu_A)_{A=1,2}$ is known, the Sachs basis is entirely determined by the angle $\vartheta$.

In order to determine the evolution of this angle, it is convenient to
rewrite Eq.~\eqref{eq:Fermi_Walker_Bianchi} in terms of the components
of $s_A$ over a tetrad basis $(e_a)_{a=1\ldots3}$ rather than over the
coordinate basis $(\partial_i)_{i=1\ldots 3}$. The choice
$e_a^i=\exp(-\beta_i)\delta^i_a$ and $e^a_i=\exp(\beta_i)\delta_i^a$ implies that the components $s_A^a\define g(s_A,e_a)$ read
\begin{equation}
(s_A^a)'+d^a(d_b)'s_A^b = 0,
\end{equation}
thus
\begin{equation}
(\cos\vartheta)'=(n_{1a} s_1^a)'= (n_{1a})'s_1^a - \underbrace{n_{1a} d^a}_{=0} (d_b)'s_1^b.
\end{equation}
Since $n_1$ is normalized, $(n_{1a})'n_1^a=0$, so $(n_{1a})'=(n_{1b})'n_2^b n_{2a} + (n_{1b})'d^b d_a$, therefore
\begin{equation}
(\cos\vartheta)'  = (n_{1b})'n_2^b n_{2a} s_1^a = (n_{1b})'n_2^b \sin\vartheta,
\end{equation}
which finally reduces to
\begin{equation}\label{thetaprime}
\vartheta' = -(n_{1a})'n_2^a = (n_{2a})'n_1^a.
\end{equation}
Summarizing, if a basis $(n_1^\mu,n_2^\mu)$ can be found, then the Sachs basis is completely determined by Eq.~(\ref{eq:s1}) with $\vartheta$ given by the integral of $(n_{2a})'n_1^a$.

\subsection{Evolution matrix}\label{subsec:evolution_matrix}

Let $\vect{\evolution}$ be the $2\times 2$ matrix that relates
the components $s_A^i$ of the Sachs basis to their values at $O$, $(s_A^i)\obs\define s_A^i(\eta\obs)$,
\begin{equation}
s_A^i(\eta) = \evolution_{AB}(\eta\leftarrow\eta\obs) (s_B^i)\obs.
\label{eq:definition_evolution}
\end{equation}
It is straightforward to show that this \emph{evolution matrix} is the solution of
\begin{align}
\evolution_{AB}' + \sigma_{AC} \evolution_{CB} &= 0,
\label{eq:evolution}\\
\evolution_{AB}(\eta\obs\leftarrow\eta\obs) &=\delta_{AB},\label{eq:evolution611}
\end{align}
where $\sigma_{AB}\define s_A^i s_B^j \sigma_{ij}$. Note that, by definition \eqref{eq:definition_evolution},
\begin{equation}
\evolution_{AB}(\eta\leftarrow\eta\obs) = s_A^i(\eta) s_{Bi}(\eta\obs).
\label{eq:evolution_expression}
\end{equation}
Note also that the position of $i$ \emph{does matter} in the above relation, because the vectors $s_A(\eta)$ and $s_A(\eta\obs)$ do not live in the same tangent spaces of the spacetime manifold~$\mathcal{M}$. The former live in $\mathrm{T}_\eta(\mathcal{M})$, their indices are raised and lowered by $\gamma_{ij}(\eta)$, while the latter live in $\mathrm{T}_{\eta\obs}(\mathcal{M})$, their indices are raised and lowered by $\gamma_{ij}(\eta\obs)=\delta_{ij}$.

In fact, inverting the position of the $i$ indices in Eq.~\eqref{eq:evolution_expression} leads to the \emph{transposed inverse}~$(\vect{\evolution}^{-1})^T$ of the evolution matrix, because
\begin{align}
s_{Ai}(\eta) s_B^i (\eta\obs) &= s_{Ai}(\eta) \evolution_{BC}(\eta\obs\leftarrow\eta) s_C^i (\eta)\nonumber\\
												&= \evolution_{BA}(\eta\obs\leftarrow\eta)\nonumber\\
												&= \evolution^{-1}_{BA}(\eta\leftarrow\eta\obs).
\end{align}
It is straightforward to check that $(\vect{\evolution}^{-1})^T$ satisfies a differential equation almost identical to Eq.~\eqref{eq:evolution}, \emph{except for a minus sign} before $\sigma_{AC}$,
\begin{equation}
(\evolution^{-1}_{BA})' - \sigma_{AC} \evolution^{-1}_{BC} = 0.
\label{eq:transposed_inverse_evolution}
\end{equation}

Using the general solution for the Sachs basis constructed in Sec.~\ref{subsec:transport_Sachs}, the evolution matrix and its transposed inverse take the form
\begin{align}
\vect{\evolution}
&= 
\begin{bmatrix}
\cos\vartheta & \sin\vartheta \\
-\sin\vartheta & \cos\vartheta
\end{bmatrix}
\begin{bmatrix}
n_1^i (s_{1i})\obs & n_1^i (s_{2i})\obs \\
n_2^i (s_{1i})\obs & n_2^i (s_{2i})\obs
\end{bmatrix},
\\
(\vect{\evolution}^{-1})^T
&=
\begin{bmatrix}
\cos\vartheta & \sin\vartheta \\
-\sin\vartheta & \cos\vartheta
\end{bmatrix}
\begin{bmatrix}
n_{1i} (s_1^i)\obs & n_{1i} (s_2^i)\obs \\
n_{2i} (s_1^i)\obs & n_{2i} (s_2^i)\obs
\end{bmatrix},
\end{align}
with the angle $\vartheta$ given by Eq.~\eqref{eq:angle_Sachs}.

Let us close this subsection by showing that the determinant of $\vect{\evolution}$ has a remarkably simple expression. Indeed
\begin{align}
(\det\vect{\evolution})' &= \tr(\vect{\evolution}^{-1}\vect{\evolution}') \det\vect{\evolution}\nonumber \\
										&= -\tr(\vect{\evolution}^{-1}\vect{\sigma}\vect{\evolution}) \det\vect{\evolution} \nonumber \\
										&= -\tr\vect{\sigma} \det\vect{\evolution},
\end{align}
where $\vect{\sigma}\define(\sigma_{AB})$ is the projection of $\sigma_{ij}$ on the Sachs basis, as defined below Eq.~(\ref{eq:evolution611}); its trace reads
\begin{equation}
\tr\vect{\sigma} = \sigma^A_A
							= \sigma_{ij} S^{ij}
							= \sigma^i_i - \sigma_{ij} d^i d^j,
\end{equation}
but, on the one hand, remember that Eq.~(\ref{e.contrainte}) implies
\begin{equation}
\sigma^i_i = \sum_{i=1}^3 \beta_i' = 0,
\end{equation}
and, on the other hand,
\begin{equation}
\sigma_{ij} d^i d^j = \frac{\sigma^{ij} k_i k_j }{\omegac^2} = \frac{(-\gamma^{ij} k_i k_j)'}{2\omegac^2} = -\frac{\omegac'}{\omegac},
\label{eq:sigma_dd}
\end{equation}
so that finally
\begin{equation}\label{e.detE}
(\det\vect{\evolution})' = -\frac{\omegac'}{\omegac}\det\vect{\evolution}
\qquad \text{whence} \qquad
\det\vect{\evolution} = \frac{1}{\omegac}.
\end{equation}

We shall see in Sec.~\ref{sec:Jacobi matrix} that the evolution matrix is a key ingredient in the expression of the Jacobi matrix.

\subsection{Explicit examples}\label{subsec:explicit_examples}

This subsection provides three explicit examples of orthonormal basis $(n_1,n_2)$ which can be used for the construction described in Sec.~\ref{subsec:transport_Sachs}, and the associated rotation angle~$\vartheta$. For the last example, we also give the expression of the evolution matrix.

\subsubsection{Frenet basis}

Since $d^\mu$ is a unit vector, it is easy to construct a vector orthogonal to it from its own derivative. Here again, calculations are easier if one works with the components over the tetrad basis $(e_a)_{a=1\ldots 3}$. We thus define
\begin{equation}
n_1^a \define \frac{(d^a)'}{\sqrt{(d_b)'(d^b)'}},
\end{equation}
and complete it by $n_2^a \define \eps\indices{^a_b_c} d^b n_1^c$. In terms of components over the coordinate basis $(\partial_i)$, we have
\begin{align}
n_1^i &= \pa{ \sigma^i_j d^j + \frac{\omegac'}{\omegac} d^i } \pa{d^i \sigma_{ij} S^j_k \sigma^k_\ell d^\ell}^{-1/2},
\label{eq:n1_example}\\
n_2^i &= \eps\indices{^i_j_k} d^j n_1^k.
\label{eq:n2_example}
\end{align}

The equation~(\ref{thetaprime}) for the evolution angle $\vartheta$ then reads
\begin{equation}
\vartheta'
= (n_{2a})'n_1^a
= \frac{\eps_{abc} d^a (d^b)' (d^c)''}{(d^a)'(d_a)'},
\end{equation}
which, in terms of components over the coordinate basis, becomes
\begin{equation}
\vartheta' = \frac{ \eps^{ijk} d_i \beta'_j d_j \pac{(\beta'_k)^2 d_k - \beta''_k d_k} }
																					{ d^i \sigma_{ij} S^j_k \sigma^k_\ell d^\ell }.
\label{eq:angle_Sachs}
\end{equation}
Interestingly, the two terms in the numerator of Eq.~\eqref{eq:angle_Sachs} are sourced by distinct geometrical properties of the Bianchi I spacetime. On the one hand, the term in $(\beta'_k)^2$ is essentially a Vandermonde determinant,
\begin{equation}
\eps^{ijk} d_i \beta'_j d_j (\beta'_k)^2 d_k
=
d_1 d_2 d_3 \prod_{i>j} (\beta'_i-\beta'_j).
\end{equation}
It depends on the \emph{triaxiality} of the Bianchi spacetime, and vanishes for an axisymmetric Bianchi I since two $\beta'_i$ are equal. On the other hand, the term in $\beta_k''$ in Eq.~\eqref{eq:angle_Sachs} can be rewritten in terms of matter's anisotropic stress. Indeed, using Eq.~\eqref{eq:dynamics_shear} and $\sigma^j_i=\beta'_i\delta^j_i$ (without summation), we get
\begin{equation}
\eps^{ijk} d_i \beta'_j d_j \beta''_k d_k = 8\pi G a^2 \eps^{ijk} d_i \sigma^\ell_j d_\ell \pi^m_k d_m.
\end{equation}
Thus, with the choice of Eqs.~(\ref{eq:n1_example}--\ref{eq:n2_example}) for $(n_1,n_2)$, the angle $\vartheta$ is ruled by an equation of the form
\begin{equation}
\vartheta' = \vartheta'\e{tri} + \vartheta'\e{stress},
\end{equation}
where $\vartheta'\e{tri}$ and $\vartheta'\e{stress}$ vanish in, respectively, an axisymmetric and anisotropic-stress-free Bianchi I model.

Though having interesting properties, the Frenet basis presented in this paragraph suffers from singularities: for a beam propagating along a principal axis of the Bianchi spacetime, $d^a=\mathrm{cst}$, so that $n_1$ cannot be defined. The next two examples will be free from such problems.

\subsubsection{Initial basis}

Another way of constructing vectors which keep orthogonal to $d^\mu$ is to use that $k_i$ are constants of motion [see Eq.~(\ref{e41})], which implies that the covariant vector $d_i=\omegac^{-1} k_i$ always points towards the same direction. Thus, the Sachs basis~$(s_A^i)\obs$ at $O$ remains orthogonal to $d_i(\eta)$ at any time:
\begin{equation}
\forall \eta \qquad d_i(\eta) (s_A^i)\obs = 0.
\end{equation}
This motivates the following definitions,
\begin{align}\label{eq:n1_example2}
\left\{
\begin{aligned}
n_1^i &\define  \frac{(s_1^i)\obs}{\sqrt{\mathring{\gamma}_{11}}}\\
n_2^i &\define \eps\indices{^i_j_k} d^j n_1^k,
\end{aligned}
\right.
\end{align}
with $\mathring{\gamma}_{11}(\eta)\define\gamma_{ij}(\eta)(s_1^i s_1^j)\obs$. Note that $n_2^i$ cannot be constructed from $(s_2^i)\obs$ in the same way as $n_1^i$ is from $(s_1^i)\obs$, because then $n_1$ and $n_2$ would not be orthogonal to each other.

In this example, the angle~$\vartheta$ reads
\begin{equation}
\vartheta' = (n_{2a})'n_1^a 
				= -\sigma_{ij} n_1^i n_2^j
				= \frac{-\eps_{ijk} d^i \sigma^k_\ell (s_1^j s_1^\ell)\obs}{\gamma_{ij}(s_1^is_1^j)\obs}.
\end{equation}
All these quantities are well behaved, as long as $\mathring{\gamma}_{11}\not=0$.

\subsubsection{Symmetrized initial basis}
\label{subsubsec:best_example}

The construction of the previous example can be slightly improved in order to be more symmetric. As mentioned above, if we define
\begin{equation}
v_1^i \define  \frac{(s_1^i)\obs}{\sqrt{\mathring{\gamma}_{11}}},
\qquad
v_2^i \define  \frac{(s_2^i)\obs}{\sqrt{\mathring{\gamma}_{22}}},
\end{equation}
with
\begin{equation}
\mathring{\gamma}_{AB}(\eta) \define \gamma_{ij}(\eta) (s_A^i s_B^j)\obs
\end{equation}
as in Eq.~\eqref{eq:n1_example2}, then $v_A^i$ is normalized and $d_i v_A^i=0$, but $v_1$ and $v_2$ are not orthogonal to each other. Let us call $\delta(\eta)$ the angle expressing their departure from orthogonality,
\begin{equation}
\cos\pa{\frac{\pi}{2}+\delta}
=
-\sin\delta
\define
\gamma_{ij} v_1^i v_2^j = \frac{\mathring{\gamma}_{12}}{\sqrt{\mathring{\gamma}_{11}\mathring{\gamma}_{22}}}.
\end{equation}
Albeit not orthogonal itself, $(v_1,v_2)$ can easily be used to obtain an orthonormal basis. Like for any couple of unit vectors, $v_1+v_2$ is orthogonal to $v_1-v_2$, which encourages us to define
\begin{equation}
n_\pm^i \define \frac{v_1^i \pm v_2^i}{\sqrt{2\mp 2\sin\delta}}.
\end{equation}
This could be used as the orthonormal basis of this last example, however we will prefer its rotation by $\pi/4$,
\begin{equation}
n_1^i \define \frac{1}{\sqrt{2}}\pa{n_+^i + n_-^i},
\qquad
n_2^i \define \frac{1}{\sqrt{2}}\pa{n_+^i - n_-^i},
\end{equation}
so that $(n_A^i)\obs=(s_A^i)\obs$, i.e. $\vartheta\obs=0$. In this case, and after a few calculations, we obtain that the angle~$\vartheta$ reads
\begin{align}
\vartheta' &= (n_{2a})'n_1^a = (n_{+a})'n_-^a \\
				&= \frac{1}{4} \tan\delta\pac{ \ln\pa{\frac{\mathring{\gamma}_{22}}{\mathring{\gamma}_{11}}} }',
\end{align}
that can also be written $(\tan\delta) \sigma_{ij}(v_2^i v_2^j -v_1^i v_1^j)/2$.

Finally, let us also give the (transposed inverse) evolution matrix which, in the present example, enjoys the relatively simple expression
\begin{multline}
(\vect{\evolution}^{-1})^T
=
\begin{bmatrix}
\cos\vartheta & \sin\vartheta \\
-\sin\vartheta & \cos\vartheta
\end{bmatrix}
\begin{bmatrix}
\cos(\delta/2) & \sin(\delta/2) \\
\sin(\delta/2) & \cos(\delta/2)
\end{bmatrix}\\
\cdot 
\begin{bmatrix}
\sqrt{\mathring{\gamma}_{11}} & 0\\
0 & \sqrt{\mathring{\gamma}_{22}}
\end{bmatrix}.
\label{eq:evolution_example3}
\end{multline}
Note that the second matrix of Eq.~\eqref{eq:evolution_example3} is \emph{not} a rotation matrix. From this result one can deduce the interesting relation
\begin{equation}
\omegac = \det \vect{\evolution}^{-1} 
= \sqrt{\mathring{\gamma}_{11}\mathring{\gamma}_{22}} \cos\delta
= \sqrt{\mathring{\gamma}_{11}\mathring{\gamma}_{22}
			-\mathring{\gamma}_{12}^2},
\end{equation}
which can also be checked by brute-force calculation.

\section{Jacobi matrix in a conformal Bianchi I geometry}
\label{sec:Jacobi matrix}

As in the previous one, all the calculations of this section are performed in the conformal geometry $\conformal{g}_{\mu\nu}$. However, all the tildes will here be carefully written, because nonintermediary results are derived.

\subsection{General solution for the Jacobi matrix}

Let us now solve the Jacobi matrix equation
\begin{equation}\label{e71}
\ddf[2]{\jacobic_{AB}}{\conformal{v}} = \opticalc_{AC} \jacobic_{CB},
\end{equation}
where we recall that the optical tidal matrix is defined by
\begin{equation}
\opticalc_{AB} \define -\conformal{R}_{\mu\nu\rho\sigma} \conformal{k}^\mu \conformal{s}_A^\nu \conformal{k}^\rho \conformal{s}_B^\sigma.
\end{equation}
The nonzero components of the Riemann tensor for the conformal Bianchi I geometry being
\begin{align}
\conformal{R}_{0i0j} = \sigma_i^k \sigma_{kj} - \sigma'_{ij},\qquad
\conformal{R}_{ijk\ell} = 2 \sigma_{k[i}\sigma_{j]\ell}.
\end{align}
A straightforward calculation, using in particular Eqs.~\eqref{eq:Fermi_Walker_Bianchi} and \eqref{eq:sigma_dd}, then leads to
\begin{equation}
\opticalc_{AB} = \omegac^2 \pac{ (\sigma_{AB})' + \sigma_{AC}\sigma_{CB} + \frac{\omegac'}{\omegac} \sigma_{AB} }.
\label{eq:optical_tidal_matrix}
\end{equation}
Therefore, since $\dd/\dd \conformal{v} = \omegac \dd/\dd\eta$, Eq.~(\ref{e71}) reads
\begin{equation}
\jacobic_{AB}''+ \frac{\omegac'}{\omegac} \jacobic_{AB}' 
= \pac{ (\sigma_{AC})' + \sigma_{AD}\sigma_{DC} + \frac{\omegac'}{\omegac} \sigma_{AC} } \jacobic_{CB} .
\label{eq:Sachs_equation_explicit}
\end{equation}

Now notice that if a matrix~$M_{AB}$ is solution of $M_{AB}'=\sigma_{AC}M_{CB}$, then it is also solution of Eq.~\eqref{eq:Sachs_equation_explicit}. Comparing with Eq.~\eqref{eq:transposed_inverse_evolution}, we deduce that the transposed inverse~$(\vect{\evolution}^{-1})^T$ of the evolution matrix is such a solution. However, it is \emph{not} the Jacobi matrix, because it does not satisfy the right initial conditions~\eqref{eq:initial_Jacobi_1} and \eqref{eq:initial_Jacobi_2}, but rather
\begin{align}
(\evolution^{-1})_{BA}(\eta\obs\leftarrow\eta\obs) &= \delta_{AB},\\
\ddf{(\evolution^{-1})^T_{AB}}{v}(\eta\obs\leftarrow\eta\obs) &= (\sigma_{AB})\obs.
\end{align}
From this particular solution, one can obtain the Jacobi matrix by use, for instance, of the method of the ``variation of the constant'' to get
\begin{equation}\label{eq:Jacobi_abtract}
\vect{\jacobic}(\eta\source\leftarrow\eta\obs) 
= (\vect{\evolution}^{-1})^T
	\int_{\eta\source}^{\eta\obs} \omegac^{-1} \vect{\evolution}^T
        \vect{\evolution} \: \dd\eta.
\end{equation}
This formula is the main result of our article. Since
\begin{align}
\evolution_{AB}(\eta\source\leftarrow\eta\obs) &= \conformal{s}_A^i(\eta\source) \conformal{s}_{Bi}(\eta\obs),\\
(\evolution^{-1})^T_{AB}(\eta\source\leftarrow\eta\obs) &= \conformal{s}_{Ai}(\eta\source) \conformal{s}_B^i(\eta\obs),
\end{align}
it can also be rewritten in terms of the components of the Sachs basis as 
\begin{multline}\label{eq:Jacobi_components}
\jacobic_{AB}(\eta\source\leftarrow\eta\obs) = (\conformal{s}_{Ai})\source (\conformal{s}_C^i\conformal{s}_{Cj})\obs \\
													\times \pa{\int_{\eta\source}^{\eta\obs} \omegac^{-1} \conformal{S}^{jk} \;\dd\eta } (\conformal{s}_{Bk})\obs.
\end{multline}
This form of the Jacobi matrix, entirely determined by the Sachs basis, reminds us about the recent results of Refs.~\cite{fanizza_lensing_2014,fanizza_exact_2013}, based on the geodesic-light-cone coordinates~\cite{2011JCAP...07..008G}. The connection between the two formalisms is left for further studies.

\subsection{An explicit expression}
\label{subsec:explicit_expression_Jacobi}

Of course, Eq.~\eqref{eq:Jacobi_abtract} cannot be considered explicit as long as one does not have an expression for $\vect{\evolution}$, which was precisely the purpose of Sec.~\ref{sec:Sachs_basis}. Here, we choose to use the results of our third example (Sec.~\ref{subsubsec:best_example}): plugging the expression~\eqref{eq:evolution_example3} of $\vect{\evolution}$ into Eq.~\eqref{eq:Jacobi_abtract}, we obtain
\begin{multline}
\vect{\jacobic}(\eta\source\leftarrow\eta\obs)
=
\begin{bmatrix}
\cos\vartheta\source & \sin\vartheta\source \\
-\sin\vartheta\source & \cos\vartheta\source
\end{bmatrix}
\begin{bmatrix}
\cos(\delta\source/2) & \sin(\delta\source/2) \\
\sin(\delta\source/2) & \cos(\delta\source/2)
\end{bmatrix}\\
\cdot
\begin{bmatrix}
\sqrt{\mathring{\gamma}_{11}(\eta\source)} & 0\\
0 & \sqrt{\mathring{\gamma}_{22}(\eta\source)}
\end{bmatrix}
\int_{\eta\source}^{\eta\obs}
\frac{\dd\eta}{\omegac^3}
\begin{bmatrix}
\mathring{\gamma}_{22} & \mathring{\gamma}_{12}\\
\mathring{\gamma}_{12} & \mathring{\gamma}_{11}
\end{bmatrix}
\label{eq:Jacobi_explicit}
\end{multline}
where the various quantities are defined in Sec.~\ref{subsubsec:best_example}, and $\mathrm{thing}\source\define\mathrm{thing}(\eta\source)$. In particular,
\begin{equation}
\vartheta\source
=
\frac{1}{4}\int_{\eta\source}^{\eta\obs} \frac{\mathring{\gamma}_{12}}{\omegac}
				\pac{ \ln\pa{\frac{\mathring{\gamma}_{22}}{\mathring{\gamma}_{11}}} }' \:\dd\eta.
\end{equation}

\subsection{Angular diameter distance}

The angular diameter distance is related to the Jacobi matrix via Eq.~\eqref{eq:angular_distance_Jacobi}, that is here
\begin{equation}
\conformal{D}\e{A} = \sqrt{\det \vect{\evolution}^{-1}} 
					\sqrt{\det\int_{\eta\source}^{\eta\obs} \omegac^{-1} \vect{\evolution}^T \vect{\evolution} \: \dd\eta} .
\label{eq:angular_distance_1}
\end{equation}
We have already seen at the end of Sec.~\ref{subsec:evolution_matrix} that the determinant of $\vect{\evolution}^{-1}$ is $\omegac$, see Eq.~(\ref{e.detE}), so that
\begin{equation}
\conformal{D}\e{A} = \sqrt{ \omegac \Delta },
\label{eq:angular_distance_2}
\end{equation}
where $\Delta$ denotes the second determinant involved in Eq.~\eqref{eq:angular_distance_1}. As originally found by Saunders in Ref.~\cite{saunders_observations_1969}, this determinant admits the remarkably simple expression
\begin{equation}\label{eq:Delta_Saunders}
\Delta = \sum_{i\not=j\not=\ell} I_i I_j k_\ell^2
\end{equation}
with
\begin{equation}
I_i \define \int_{\eta\source}^{\eta\obs} \omegac^{-3} \ex{2\beta_i} \:\dd\eta.
\label{eq:integral_Saunders}
\end{equation}
It is however surprising that the author of Ref.~\cite{saunders_observations_1969} gives this nontrivial expression of $\Delta$ with no derivation. Since we did not find any elsewhere in the literature, we propose one in the Appendix. 
Note that, by computing directly the determinant of the explicit expression~\eqref{eq:Jacobi_explicit}, one can obtain an alternative form---though mathematically equivalent---of Saunders' determinant
\begin{equation}
\Delta = \mathcal{I}_{11} \mathcal{I}_{22} - \mathcal{I}_{12}^2,
\end{equation}
with
\begin{equation}
\mathcal{I}_{AB} \define \int_{\eta\source}^{\eta\obs} \omegac^{-3} \mathring{\gamma}_{AB} \:\dd\eta.
\end{equation}

\subsection{The weak shear regime}
\label{subsec:WSR}

Our solution for the Jacobi matrix is completely general, which means that it remains valid even for very anisotropic Bianchi I spacetimes [with $\beta_i=\mathcal{O}(1)$]. However, because cosmological observations suggest that our Universe is extremely close to isotropic, it can be interesting in practice to study the weak-shear behavior of our solution. We now perform such an expansion of the Jacobi matrix---and the related quantities---at first order in $\beta_i \ll 1$, in the conformal Bianchi I geometry.

In this regime, the cyclic frequency of the photons and the evolution matrix of the Sachs basis respectively read
\begin{align}
\omegac &= 1 - \B + \mathcal{O}(\beta_i^2), \\
\evolution_{AB} &= \delta_{AB} + \B_{AB} + \mathcal{O}(\beta_i^2),
\end{align}
where we have defined the first order quantities
\begin{align}
\B_{AB}(\eta) &\define \sum_{i=1}^3 \beta_i(\eta) (s_A^i s_B^i)\obs,\\
\B(\eta) &\define \sum_{i=1}^3 \beta_i(\eta) k_i^2 = -\tr(\B_{AB}).
\end{align}
Note that, in terms of the notations of Sec.~\ref{subsec:explicit_examples}, $\mathring{\gamma}_{AB}=\delta_{AB}+2\B_{AB}+\mathcal{O}(\beta_i^2)$. The expression of the Jacobi matrix is then easily found to be
\begin{multline}
\jacobic_{AB}(\eta\source\leftarrow\eta\obs) = \delta_{AB} \pac{(\eta\source-\eta\obs)+\int_{\eta\source}^{\eta\obs} \B\;\dd\eta}\\
+ (\eta\source-\eta\obs)\B_{AB}(\eta\source) - 2 \int_{\eta\source}^{\eta\obs} \B_{AB}\;\dd\eta +\mathcal{O}(\beta_i^2).
\label{eq:Jacobi_WSR}
\end{multline}

Note that, at this order, the Jacobi matrix remains \emph{symmetric}. In terms of the decomposition of Eq.~\eqref{eq:decomposition_Jacobi}, it means that the rotation angle vanishes $\psi=\mathcal{O}(\beta_i^2)$. The angular diameter distance is obtained by computing the (square root of the) determinant of \eqref{eq:Jacobi_WSR}, which leads to
\begin{equation}
\conformal{D}\e{A} = \pa{1-\frac{\B\source}{2}}(\eta\obs-\eta\source) + 2 \int_{\eta\source}^{\eta\obs} \B\;\dd\eta + \mathcal{O}(\beta_i^2).
\label{eq:DA_WSR}
\end{equation}
Finally, the optical shear, encoded into the exponential matrix of Eq.~\eqref{eq:decomposition_Jacobi} is at this order equal to the traceless part of the Jacobi matrix~$\jacobic_{\langle AB \rangle}$,
\begin{multline}
\begin{bmatrix}
-\Gamma_1 & \Gamma_2 \\
\Gamma_2 & \Gamma_1
\end{bmatrix}
=
(\eta\source-\eta\obs) \B_{\langle AB \rangle}(\eta\source) \\- 2 \int_{\eta\source}^{\eta\obs} \B_{\langle AB \rangle}\;\dd \eta + \mathcal{O}(\beta_i^2),
\label{eq:shear_WSR}
\end{multline}
where $\langle A B \rangle$ means the traceless part with respect to $\delta_{AB}$; in particular $\B_{\langle AB \rangle}= \B_{AB} - \B\,\delta_{AB}/2$. Note that the above shear does not need to be tilded, because $\vect{\jacobic}\propto\vect{\jacobi}$ so that $(\conformal{\Gamma}_1,\conformal{\Gamma}_2)=(\Gamma_1,\Gamma_2)$.

\section{Summary}
\label{sec:summary}

Before concluding, let us summarize the main results of this paper, under the form of a recipe for the reader who would like to use them in practice. It is also the occasion to recover the untilded quantities from the tilded ones using the dictionary of Sec.~\ref{sec:dictionary}.
\begin{enumerate}
\item Solve for the cosmology (Sec.~\ref{sec:Bianchi_I_spacetime}) to determine the scale factor~$a(\eta)$, and the functions $\beta_i(\eta)$ characterizing the spatial conformal metric~$\gamma_{ij}$. Set by convention $a(\eta\obs)=1$ and $\beta_i(\eta\obs)$ so that $(g_{\mu\nu})\obs=\eta_{\mu\nu}$. Note that, by virtue of the dictionary of Sec.\ref{sec:dictionary}, all conformal (tilded) quantities are equal to their untilded counterpart at $\eta=\eta\obs$. An example of such dynamics can be found in Ref.~\cite{PPU2}.
\item Pick a direction of observation $d^i\obs$ on the sky and an initial Sachs basis $(s_A^i)\obs$ orthogonal to it. A possible choice using spherical coordinates~$(\theta\obs,\ph\obs)$ is
\begin{align}
(d^i)\obs &= (\sin\theta\obs\cos\ph\obs,\sin\theta\obs\sin\ph\obs,\cos\theta\obs)\\
(s_1^i)\obs &= (\cos\theta\obs\cos\ph\obs,\cos\theta\obs\sin\ph\obs,-\sin\theta\obs)\\
(s_2^i)\obs &= (-\sin\ph\obs,\cos\ph\obs,0)
\end{align}
\item Set by convention $\omega\obs=1$. The wave four-vector of the photon is then characterized at any time by $k_i=\mathrm{cst}=d^i\obs$ and $k^t=\omega=\omegac/a$ where $\omegac$ is given by Eq.~\eqref{Defomegatilde}. This is enough to compute the redshift~$z\define1/\omega-1$, the redshift drift~\eqref{eq:redshift_drift}, the direction drift~\eqref{eq:direction_drift}, and the angular diameter distance $D\e{A}=a\conformal{D}\e{A}=\sqrt{a^3\omega\Delta}$, where $\Delta$ is given by Eqs.~\eqref{eq:Delta_Saunders} and \eqref{eq:integral_Saunders}). In the weak shear regime, use the expression~$\eqref{eq:DA_WSR}$ for $\conformal{D}\e{A}$.
\item In order to get the full Jacobi matrix~$\vect{\jacobi}$, first determine the evolution matrix~$\vect{\evolution}$ using the method described in Sec.~\ref{sec:Sachs_basis}, then plug it into Eq.~\eqref{eq:Jacobi_abtract} to obtain $\vect{\jacobic}$. An example of this procedure had been given in Sec.~\ref{subsec:explicit_expression_Jacobi}. Apply finally the conformal dictionary relation $\vect{\jacobi}=a\vect{\jacobic}$.
\item Quantities such as optical shear and optical rotation are obtained by performing the canonical decomposition~\eqref{eq:decomposition_Jacobi} of the obtained Jacobi matrix. Their weak-shear expressions are the ones obtained in Sec.~\ref{subsec:WSR}.
\end{enumerate}

\section{Conclusion}
\label{sec:conclusion}

This article detailed an analytic integration of all the equations governing light propagation in a Bianchi I spacetime. From a technical point of view, the symmetries of the problem were central in our derivations. First, in Sec.~\ref{sec:light_rays_BIanchi_I}, the invariance of the metric under spatial translation allowed us to solve the null geodesic equation without any calculation. Second, the invariance of the equations governing light propagation under conformal transformations allowed us to greatly simplify the calculation of the Jacobi matrix in Sec.~\ref{sec:Jacobi matrix}.

As a first output, we obtained formulas for the redshift and direction drift in a Bianchi I universe, which are comparable to former papers generally restricted to Lema\^{\i}tre-Tolman-Bondi spacetimes~\cite{zdotuce,realtimecosmo2}. As a second output and sanity check, we recovered the already known~\cite{saunders_observations_1968,saunders_observations_1969,schucker_bianchi_2014} expression of the angular diameter distance. However, we emphasize that our results are more powerful, because they also give access to the complete lensing behavior of Bianchi~I, including optical shear and rotation. This new step will be the starting point of a deeper analysis of light propagation in a \emph{perturbed} Bianchi~I spacetime, which would allow us to evaluate the amplitude of the comic shear $B$-mode signal associated with a violation of local isotropy, as predicted by Ref.~\cite{Bmodes}.

Our study can therefore be used to set constraints on the spatial isotropy of the Hubble flow from the analysis of the Hubble diagram, but also from possible future observation such as the redshift drift~\cite{codex,codex2} (see e.g. Ref.~\cite{realtimecosmo} for a review of the observational possibilities concerning both the time and direction drifts). Together with weak lensing~\cite{Bmodes}, this offers a set of tools to constrain any late-time anisotropy of cosmic expansion. 

\begin{acknowledgments}
  \noindent
  This work was made in the ILP LABEX (under reference ANR-10-LABX-63) and was supported
  by French state funds managed by the ANR within the Investissements
  d'Avenir programme under reference ANR-11-IDEX-0004-02 and the Programme National de Cosmologie et Galaxies.  
  \end{acknowledgments}
  
\appendix
\section*{Appendix: Derivation of Saunders' formula}
\label{app:derivation_Saunders_formula}

Let us calculate Saunders' determinant~\cite{saunders_observations_1969}, defined as
\begin{align}
\Delta &\define \det\int_{\eta\source}^{\eta\obs} \omegac^{-1} \vect{\evolution}^T \vect{\evolution} \: \dd\eta \\
			&= \det\pac{(s_{Ai})\obs \intmatrix^{ij} (s_{Bj})\obs}
\end{align}
where we have denoted
\begin{equation}
\intmatrix^{ij} \define \int_{\eta\source}^{\eta\obs} \omegac^{-1} \conformal{S}^{ij} \: \dd\eta,
\end{equation}
so that the quantity $\Delta$ is the determinant of the restriction of $\vect{\intmatrix}\define(\intmatrix^{ij})$ on the 2-plane spanned by $[(s_{Ai})\obs]_{A=1,2}$. It turns out that this restriction actually encodes the whole matrix $\vect{\intmatrix}$. Indeed, since $\intmatrix^{ij}k_i=0$ ($k_i$ being a constant, it can safely enter into the integral), it is easy to check that
\begin{equation}
\intmatrix^{ij} = \pac{ (s_{Ak})\obs \intmatrix^{k\ell} (s_{B\ell})\obs } (s_A^i s_B^j)\obs;
\end{equation}
in other words, written in the basis $[k^i,(s_1^i)\obs,(s_2^i)\obs]$, the matrix $\vect{\intmatrix}$ reads
\begin{equation}
\vect{\intmatrix}
=
\begin{bmatrix}
0 & 0 & 0 \\
0 & (s_{1i})\obs \intmatrix^{ij} (s_{1j})\obs & (s_{1i})\obs \intmatrix^{ij} (s_{2j})\obs \\
0 & (s_{2i})\obs \intmatrix^{ij} (s_{1j})\obs & (s_{2i})\obs \intmatrix^{ij} (s_{2j})\obs
\end{bmatrix}.
\end{equation}
We conclude that if $(0,\intmatrix_+,\intmatrix_-)$ denote the three eigenvalues of $\vect{\intmatrix}$, then $\Delta=\intmatrix_+\intmatrix_-$ is the product of the last two. Let us now calculate this product.

The characteristic polynomial of $\vect{\intmatrix}$ reads
\begin{align}
\chi_{\vect{\intmatrix}}(X) &\define \det(\vect{\intmatrix}-X\vect{1}_3) \\
														&= -\frac{X}{2} \pac{ (\tr\vect{\intmatrix})^2 - \tr(\vect{\intmatrix}^2)}
																+ X^2 \tr\vect{\intmatrix}
																- X^3 \\
														&=- X \intmatrix_+ \intmatrix_- + X^2 (\intmatrix_+ + \intmatrix_-) - X^3,
\end{align}
where we have used that $\det\vect{\intmatrix}=0$, and the fact that the roots of $\chi_{\vect{\intmatrix}}$ are $(0,\intmatrix_+,\intmatrix_-)$; thus
\begin{equation}
\Delta = \intmatrix_+ \intmatrix_- = \frac12 \pac{ (\tr\vect{\intmatrix})^2 - \tr(\vect{\intmatrix}^2)}.
\end{equation}
Written explicitly, the expression above is
\begin{align}
\Delta &= \intmatrix^{11}\intmatrix^{22} 
				+ \intmatrix^{11}\intmatrix^{33}+ \intmatrix^{22}\intmatrix^{33} \nonumber\\
	&\qquad			- (\intmatrix^{13})^2- (\intmatrix^{12})^2 - (\intmatrix^{23})^2,
\label{eq:Delta_complicated}
\end{align}
but it can be further simplified using again that $\intmatrix^{ij}k_i=0$, which implies
\begin{align}
\intmatrix^{11} &= - \frac{k_2}{k_1} \intmatrix^{12} - \frac{k_3}{k_1} \intmatrix^{13},\\
\intmatrix^{22} &= - \frac{k_1}{k_2} \intmatrix^{12} - \frac{k_3}{k_2} \intmatrix^{23},\\
\intmatrix^{33} &= - \frac{k_2}{k_3} \intmatrix^{23} - \frac{k_1}{k_3} \intmatrix^{13}.
\end{align}
Plugging these relations in Eq.~\eqref{eq:Delta_complicated} indeed leads to
\begin{equation}
\Delta = \frac{k_1 \intmatrix^{12} \intmatrix^{13} 
						+ k_2 \intmatrix^{12} \intmatrix^{23}
						+ k_3 \intmatrix^{13} \intmatrix^{23}}
						{k_1 k_2 k_3}.
\end{equation}
Finally, with the definitions
\begin{equation}
I_1 \define -\frac{\intmatrix^{23}}{k_2 k_3},
\quad
I_2 \define -\frac{\intmatrix^{13}}{k_1 k_3},
\quad
I_3 \define -\frac{\intmatrix^{12}}{k_1 k_2},
\label{eq:def_I_appendix}
\end{equation}
we recover Saunders' formula
\begin{equation}
\Delta = k_1^2 I_2 I_3 + k_2^2 I_1 I_3 + k_3^2 I_1 I_2.
\end{equation}
Of course, we also have to check that the $I_i$s defined in Eq.~\eqref{eq:def_I_appendix} agree with the expressions given in Eq.~\eqref{eq:integral_Saunders}. Consider for instance $I_1$, starting from
\begin{equation}
I_1 \define -\frac{\intmatrix^{23}}{k_2 k_3} = -\frac{1}{k_2 k_3} \int_{\eta\source}^{\eta\obs} \omegac^{-1} \conformal{S}^{23} \: \dd\eta.
\end{equation}
Because $\conformal{S}^{23}=\gamma^{23}-\conformal{d}^2\conformal{d}^3$, and $(\gamma^{ij})$ is diagonal, we have
\begin{align}
-\conformal{S}^{23} &= \conformal{d}^2 \conformal{d}^3 \nonumber\\
									&= \ex{-2\beta_2} \ex{-2\beta_3} \conformal{d}_2 \conformal{d}_3 \nonumber\\
									&= \ex{2\beta_1} (\omegac^{-1} \conformal{k}_2)(\omegac^{-1}\conformal{k}_3) 
									\label{eq:calculation_S23_penultimateline}\\
									&= \ex{2\beta_1} \omegac^{-2} k_2 k_3.
									\label{eq:calculation_S23_lastline}
\end{align}
whence
\begin{equation}
I_1 = \int_{\eta\source}^{\eta\obs} \omegac^{-3} \ex{2\beta_1} \: \dd\eta.
\label{eq:I1}
\end{equation}
In Eq.~\eqref{eq:calculation_S23_penultimateline} we have used that $\sum_{i=1}^3\beta_i=0$, and in Eq.~\eqref{eq:calculation_S23_lastline} the relation $\tilde{k}_i=k_i$ established in Sec.~\ref{sec:dictionary}. Equation~\eqref{eq:I1} agrees with Eq.~\eqref{eq:integral_Saunders}, and it is clear that the same calculation can be done for $I_2, I_3$.

\bibliography{bibliography}

\end{document}